# Shape anisotropy revisited in single-digit nanometer magnetic tunnel junctions

K. Watanabe[1], B. Jinnai[2], S. Fukami[1,2,3,4*], H. Sato[1,2,3,4], and H. Ohno[1,2,3,4,5]

[1]*Laboratory for Nanoelectronics and Spintronics, Research Institute of Electrical Communication, Tohoku University, 2-1-1 Katahira, Aoba-ku, Sendai 980-8577, Japan*

[2]*Center for Spintronics Integrated Systems, Tohoku University, 2-1-1 Katahira, Aoba-ku, Sendai 980-8577 Japan*

[3]*Center for Spintronics Research Network, Tohoku University, 2-1-1 Katahira, Aoba-ku, Sendai 980-8577 Japan*

[4]*Center for Innovative Integrated Electronic Systems, Tohoku University, 468-1 Aramaki Aza Aoba, Aoba-ku, Sendai 980-0845 Japan*

[5]*WPI-Advanced Institute for Materials Research, Tohoku University, 2-1-1 Katahira, Aoba-ku, Sendai 980-8577 Japan*

Nanoscale magnetic tunnel junction plays a pivotal role in magnetoresistive random access memories. Successful implementation depends on a simultaneous achievement of low switching current for the magnetization switching by spin-transfer torque and high thermal stability, along with a continuous reduction of junction size. Perpendicular-easy-axis CoFeB/MgO stacks possessing interfacial anisotropy have paved the way down to 20-nm scale, below which a new approach needs to be explored. Here we show magnetic tunnel junctions that satisfy the requirements at ultrafine scale by revisiting shape anisotropy, which is a classical part of magnetic anisotropy but has not been fully utilized in the current perpendicular systems. Magnetization switching solely driven by current is achieved for junctions smaller than 10 nm where sufficient thermal stability is provided by shape anisotropy without adopting new material systems. This work is expected to push forward the development of magnetic tunnel junctions towards single-digit-nm-scale nano-magnetics/spintronics.

Since the theoretical prediction of Spin-Transfer Torque (STT)[1,2] and early experimental demonstration of STT-induced magnetization switching[3-9], its application to writing scheme in magnetoresistive random access memories (STT-MRAMs) has been a focus of spintronics research for the last two decades[10-19]. To make it a viable technology, the magnetic tunnel junctions (MTJs), the heart of the STT-MRAMs, is required to simultaneously show high tunnel magnetoresistance (TMR) ratio, low switching current (density), and high thermal stability factor $\Delta$ ($\equiv E/k_\mathrm{B}T$, where $E$ is the energy barrier between two possible states, $k_\mathrm{B}$ the Boltzmann constant, and $T$ the absolute temperature). Furthermore, shrinking the size of the MTJ is necessary to achieve higher-density integration. The high TMR ratio was found to be attained by using MgO tunnel barrier[20,21]. Meanwhile, a lower switching current for a given $\Delta$ is known to be obtained in perpendicular-easy-axis systems compared with in-plane ones[22-25]. However, this had posed a challenge for high-performance MTJs as most of the material systems with perpendicular anisotropy do not meet all the requirements simultaneously at a satisfactory level. This issue was settled by a discovery of perpendicular-easy-axis CoFeB/MgO, which itself was a conventional in-plane system but with sufficient reduction of CoFeB thickness the buried interfacial perpendicular anisotropy[26,27] emerged as a dominant factor to achieve a perpendicular easy axis[28-30]. A double CoFeB (or FeB)/MgO interface structure was subsequently introduced, which increases the net anisotropy energy and $\Delta$ by a factor of two[31-33], allowing scaling of MTJs down to around 20 nm[34]. Using the double-interface (Co)FeB-MgO MTJ, the STT-MRAMs are about to enter the market[17]. However, when one further reduces the MTJ size, the *interfacial-anisotropy* approach reaches a physical limit in securing sufficient $\Delta$ while achieving STT-induced switching.[35,36] Thus, establishing a technology to pave the way towards sub-10 nm, or single-digit nm, MTJs is of pressing importance.

Here we show an unexplored approach to achieve perpendicular-easy-axis MTJs meeting the requirements at ultrafine scales. We employ a commonly-available material system FeB/MgO with the double-interface structure, but this time increase the thickness of FeB so that the shape anisotropy emerges as a dominant factor to keep the easy axis along the perpendicular direction. The fabricated MTJs exhibit a high $\Delta$ of more than 80, a sufficiently high value for most of the applications, and yet can be switched by STT at sizes smaller than 10 nm.

**Results**

**Shape-anistropy MTJs.** We first describe the concept of the shape-anisotropy MTJ. The shape anisotropy energy originates from a classical magnetostatic interaction, and is defined as the energy difference between states for which magnetization is fully aligned along particular directions. In general, it acts to align the magnetization along the longest direction of the samples, that is, in-plane direction for membrane samples and longitudinal direction for needle-shaped samples. Assuming single-domain magnetization reversal, $\Delta$ of MTJ being subject to the shape, bulk, and interfacial anisotropies is expressed as

$$\Delta \equiv \frac{E}{k_\mathrm{B}T} = \left(-\delta N \frac{M_\mathrm{S}^2}{2\mu_0}t + K_\mathrm{b}t + K_\mathrm{i}\right)\frac{\pi D^2}{4k_\mathrm{B}T}, \quad (1)$$

where $\mu_0$ is the permeability in free space, $K_\mathrm{b}$ and $K_\mathrm{i}$ are the bulk (magnetocrystalline) and interfacial anisotropy energy densities, respectively, and $M_\mathrm{S}$, $t$, and, $D$ are the spontaneous magnetization, thickness, and diameter of the ferromagnetic layer, respectively. $\delta N$ is the difference in dimensionless demagnetization coefficient, or the shape anisotropy coefficient, between the perpendicular and in-plane directions, which is close to 1 when $D \gg t$, as has been the case so far [see Supplementary Note 1 for more details]. Next, the intrinsic critical current of STT switching $I_\mathrm{C0}$ is given as[1,2]

$$I_\mathrm{C0} = \alpha \frac{2e\gamma}{\mu_\mathrm{B} g_\mathrm{STT}} E, \quad (2)$$

where $\alpha$ is the Gilbert damping constant, $e$ the elementary charge, $\gamma$ the gyromagnetic ratio, $\mu_\mathrm{B}$ the Bohr magneton, and $g_\mathrm{STT}$ the STT efficiency. Here, we note that the intrinsic critical voltage $V_\mathrm{C0}$ is given by the product of the intrinsic critical current density $J_\mathrm{C0} = I_\mathrm{C0}/(\pi D^2/4)$ and the resistance-area product $RA$, and should be small enough to secure reliability and integrated-circuit compatibility. Equation (1) indicates that to obtain sufficient $\Delta$ (> 60 - 80[12,36]), one needs to employ material systems with high enough $K_\mathrm{b}$ or $K_\mathrm{i}$ and low enough $M_\mathrm{S}$ to overcome the negative shape anisotropy term. Meanwhile, equation (2) indicates that $I_\mathrm{C0}$ is proportional to a product of $\alpha$ and $E$ and is not dependent on which term in the right-hand side of equation (1) dominates $E$. Thus, to achieve low $I_\mathrm{C0}$ for a given $\Delta$, one needs to employ low $\alpha$ materials. The *interfacial-anisotropy* MTJ with double-interface structure[32] satisfies the requirements on both $\Delta$ and $I_\mathrm{C0}$ down to around $D = 20$ nm[34]. However, further reduction of $D$ inevitably degrades $\Delta$, requiring a new approach. This leads us to revisit the shape anisotropy. If one can engineer δ$N$ in equation (1), the perspective is expected to drastically change.

Now, let us look at how $\Delta$ and $V_{C0}$ vary with $D$ and $t$ through the change in δ$N$. Figures 1a and 1b show $D$ and $t$ dependence of $\Delta$ and $V_{C0}$ calculated from equations (1) and (2). Material parameters are set based on a typical double-interface MgO/FeB/MgO system [see Method]. δ$N$ is calculated analytically under several approximations [see Supplementary Note 1]. The high-$\Delta$ region at bottom-right corner in Fig. 1a corresponds to the conventional *interfacial-anisotropy* MTJ. Here, $\Delta \geq 80$ cannot be obtained when $D \lesssim 15$ nm. However, when one further increases $t$, the easy axis turns to in-plane once and then becomes perpendicular again, entering a new region, that is, the *shape-anisotropy* MTJ. Here, a region with sufficient $\Delta$ is seen even at 10 nm or less by virtue of the shape anisotropy. As for $V_{C0}$ (Fig. 1b), it increases with $\Delta$ and $1/D^2$. Figure 1c shows design windows, in which necessary conditions of $\Delta \geq 80$ (red-hatched) and $V_{C0} \leq 0.5$ V (blue-hatched) are satisfied, where $RA = 1$ Ω μm$^2$ is assumed[37]. *Shape-anisotropy* approach offers a region satisfying the two requirements simultaneously at smaller $D$ than for *interfacial-anisotropy* approach, holding promise for ultrafine STT-MRAMs.

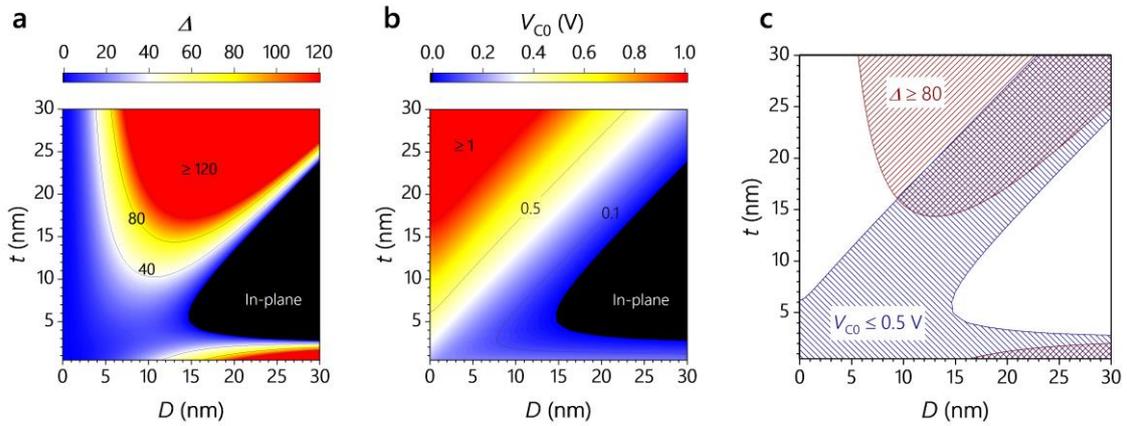

**Figure 1 | Calculation on MTJ performances. a**, Thermal stability factor $\Delta$ as a function of thickness $t$ and diameter $D$ **b**, Intrinsic critical voltage $V_{C0}$ based on the Slonczewski model. **c**, Regions satisfying $\Delta \geq 80$ (red-hatched) and $V_{C0} \leq 0.5$ V (blue-hatched), where resistance-area product $RA = 1$ Ω μm$^2$ is assumed. The regions in which the hatches are overlapped correspond to the design windows.

**Characterization of blanket film.** Following the calculation results described above, we start the experimental parts from a blanket film study. Based on the calculation, we deposit 15-nm-thick ferromagnetic layer for the shape-anisotropy MTJ. The stack structure of MTJ is thermally-oxidized Si substrate/ Ta(5)/ Ru(10)/ Ta(15)/ Pt(5)/ [Co(0.4)/ Pt(0.4)]$_{\times 6}$/ Co(0.4)/ Ru(0.4)/ Co(0.4)/ [Pt(0.4)/ Co(0.4)]$_{\times 2}$/ Ta(0.2)/ CoFeB(1)/ MgO(0.93)/ FeB(15)/ MgO(0.90)/ Ru(5) (nominal thickness in nm), as depicted in Fig. 2a [see Method for details]. FeB is employed as a material for the recording layer because it is expected to exhibit large $K_i$ and small $\alpha$. Note that large interfacial anisotropy helps obtaining perpendicular easy axis. Magnetization $M$ responses to magnetic field $H$ applied along the in-plane and out-of-plane direction ($M$-$H$ curves) for a stack consisting of only the recording layer part (MgO/ FeB/ MgO) are shown in Fig. 2b. In-plane easy axis is confirmed from the two curves. The field to saturate $M$ along the out-of-plane (hard-axis) direction, is about 1.5 T, which is close to the saturated value of $M$ (= 1.52 ± 0.01 T), indicating that the shape anisotropy is dominant in laying magnetization in the plane. $K_i$ and $K_b$ are evaluated to be 2.2 ± 0.1 mJ m$^{-2}$ and (− 1.10 ± 0.07) × 10$^5$ J m$^{-3}$ from the thickness dependence of $M$-$H$ curves. Ferromagnetic resonance (FMR) reveals that $\alpha$ of the recording layer is as small as 0.00425 ± 0.00003 [see Supplementary Note 2].

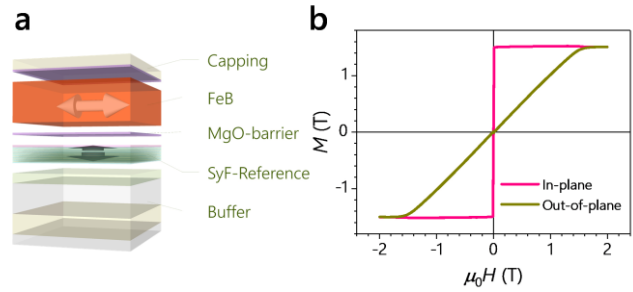

**Figure 2 | Stack structure and *M-H* curves. a**, Schematic of the blanket film of the stack structure, where the arrows show the expected magnetic easy axes. **b**, Magnetization *M* in response to external magnetic field *H* applied along the in-plane and out-of-plane directions (*R-H* curves), for a blanket film comprising the recording layer part, MgO/FeB/MgO.

**Characterization of nano MTJs.** Now we move onto the study of nano MTJs. Figure 3a shows a schematic illustration of the processed MTJ [see Method for the detail of integration process], and Figs. 3b and 3c respectively show the corresponding cross-sectional high-angle annular dark-field scanning transmission electron microscopy image and element mapping using electron energy-loss spectroscopy of a patterned MTJ. They reveal that $D$ and $t$ of the recording layer consisting of Fe and B are about 10 and 15 nm, respectively, as designed, within the region of perpendicular easy axis with high $\Delta$ [see Fig. 1a].

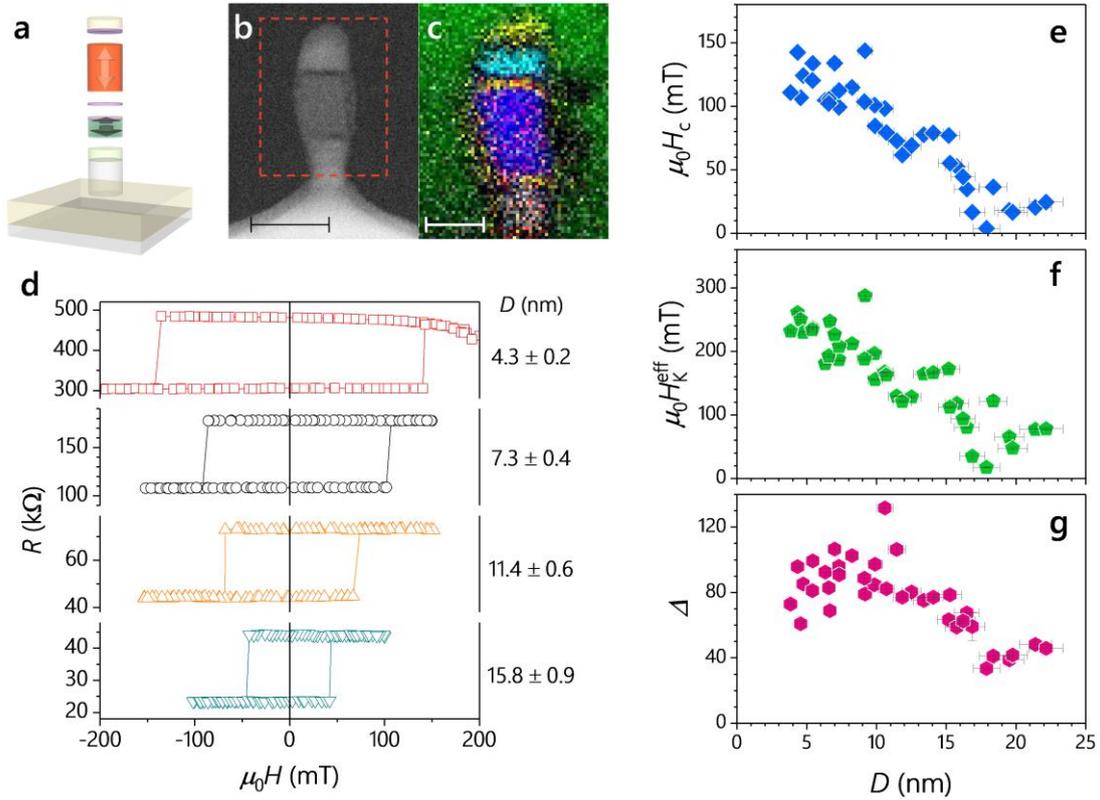

**Figure 3 | Magnetic-field characterization of nano-MTJs. a**, Schematic image of the nano MTJ, where the arrows show the expected magnetic easy axes. **b**, Cross-sectional high-angle annular dark-field scanning transmission electron microscopy images of a MTJ after the ion milling process. Broken rectangle indicates the area for element mapping shown in **c**. The scale bar corresponds to 20 nm. **c**, Corresponding image of element mapping using electron energy-loss spectroscopy. Representations of each color are B: red, N: green, O: yellow, Fe: dark blue, Co: white, Ru: light blue. The scale bar corresponds to 10 nm. **d**, MTJ resistance $R$ in response to perpendicular field $H$ ($R$-$H$ loops) for MTJs with various diameters $D$. **e**-**g**, MTJ properties as a function of $D$. **e**, Coercive field $H_C$, **f**, effective magnetic anisotropy field $H_K^{eff}$ and **g**, thermal stability factor $\Delta$. The error bars represent s.e.m, where those along the horizontal direction originate from uncertainty in

determination of resistance-area product and those along the vertical direction originate from the fitting.

Figure 3d shows the MTJ resistance $R$ as a function of the out-of-plane field ($R$-$H$ curves) for junctions with various $D$, where $D$ of each MTJ is electrically determined from its $R$ and $RA$ (= 4.5 ± 0.5 Ω μm$^2$) [see Supplementary Note 3 for detail]. Square hysteresis loops are observed, indicating perpendicular easy axis owing to the shape anisotropy. Coercive field $H_C$ increases as $D$ decreases [see Fig. 3e], also providing evidence that the shape anisotropy governs the stability of magnetization along the perpendicular direction. Next, we evaluate the effective anisotropy field $H_K^{eff}$ and $\Delta$ from switching probability using pulse magnetic field with 1 s duration [see Supplementary Note 4]. Figures 3f and 3g show the obtained $H_K^{eff}$ and $\Delta$ as a function of $D$ for a number of MTJs. $H_K^{eff}$ increases as $D$ decreases, consistent with the trend of $H_C$. Importantly, a sufficiently high $\Delta$ of more than 80 is obtained for MTJs with less than 10 nm in diameter and the values are consistent with our analytical calculation considering the shape anisotropy [see Fig. 1a]. For the smallest MTJ we measure ($D$ = 3.8 ± 0.2 nm), $\mu_0 H_C$ = 111 mT, $\mu_0 H_K^{eff}$ = 232 ± 3 mT, and $\Delta$ = 73 ± 1 are obtained. Deviation of observed trend between $\Delta$ and $D$ from that expected from calculation, particularly a larger $\Delta$ and its smaller decreasing rate with $D$ than that expected for the range of $D$ less than around 15 nm, will be discussed later.

**Current-induced magnetization switching.** Finally, we show the switching of the MTJs with small $D$ and high $\Delta$ using STT. Figure 4 shows typical $R$ versus applied current density $J$ ($R$-$J$) curves at zero magnetic fields. The pulse duration $\tau_p$ is 10 ms and $R$ is measured during the pulse application. Inset shows the $R$-$H$ curves for the same MTJ. Bidirectional magnetization switching is observed for each MTJ. In line with equation (2), the switching current density $J_{SW}$ increases as $D$ decreases due to an increase in $E/D^2$. The smallest MTJ in which the switching is observed is 8.8 ± 0.5 nm in diameter. STT-induced switching for even smaller MTJs is expected to be achieved by reducing $t$ and/or $RA$. From the observed $J_{SW}$, $J_{C0}$ can be calculated by[38]

$$J_{C0} = \frac{J_{SW}}{1 - \frac{1}{\Delta}\ln\left(\frac{\tau_p}{\tau_0}\right)}, \qquad (3)$$

where $\tau_0$ is the inverse of attempt frequency (= 1 ns). For the case with 10.4-nm MTJ, where $J_{SW}$ ≈ 3.2 × 10$^{11}$ A m$^{-2}$ and $\Delta$ ≈ 90, $J_{C0}$ is derived to be 3.9 × 10$^{11}$ A m$^{-2}$, corresponding to the intrinsic critical current $I_{C0}$ of 42 μA. This leads to the switching efficiency, defined as $\Delta/I_{C0}$, of 2.1 μA$^{-1}$. This efficiency is almost the same as that in previously-reported *interfacial-anisotropy* MTJs with similar size, which exhibited a much reduced $\Delta$[30,34]. Meanwhile, the theoretical value calculated from equation (2) is $4.4 \times 10^{11}$ A m$^{-2}$. Good correspondence indicates that the observed current-induced switching is well described by the single-domain model considering the shape anisotropy, despite a large thickness of the recording layer.

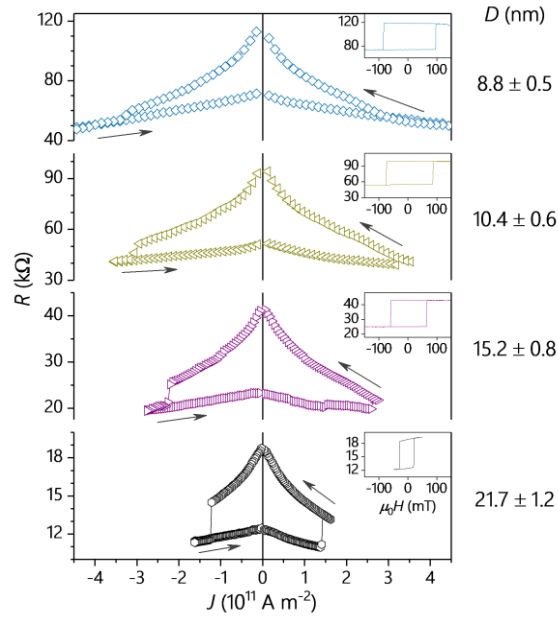

**Figure 4 | Current-induced switching properties.** *R* in response to applied current density *J* (*R-J* loops) for MTJs with various diameters *D*. The insets show the *R-H* loop for the same MTJ.

**Discussion**

As presented above, high-$\Delta$ ($\geq 80$) MTJs that can be switched by current are obtained at around and below 10 nm. The obtained $\Delta$ and $J_{SW}$ are explained by an analytical model (equations (1) and (2)), which considers the shape anisotropy. Here, we discuss several other factors that may be necessary to be included to further describe the experiment. At first, while the values of material parameters such as $M_S$, $K_i$, $K_b$, and $\alpha$ used for the calculation are determined based on blanket film studies, they could be different between the blanket film and nano MTJs; moreover they could vary with *D*. For instance, our recording layer is surrounded by a Si-N passivation layer, which could absorb B in the FeB recording layer and also could azotize the FeB recording layer. Such chemical reactions may result in variation of magnetic parameters. Secondly, the approximation used to calculate δ$N$ [see Supplementary Note 1] may result in a certain deviation. Thirdly, as can be seen in *R-J* loops shown in Fig. 4, TMR ratio has a bias-voltage dependence, resulting in overestimation of $g_{STT}$ in equation (2). Finally, incoherent magnetization behavior and spin relaxation along the perpendicular direction may arise due to the thick recording layer, and there could be an upper limit on the thickness in which the present model holds true. Thus, to better understand the *shape-anisotropy* MTJs, the issues raised above have to be elucidated in future.

$J_{SW}$, which is observed to be of the order of $10^{11}$ A/m$^2$, has to be reduced for practical use in terms of drivability of cell transistors, active power consumption, reliability issues, and so on. This is particularly important when one considers further scaling of MTJ beyond 8.8 nm, since the theoretical model [equation (2)] predicts further increase in $J_{SW}$ with decreasing *D* if *Δ* is kept at a certain level. The reduction of $J_{SW}$ is expected to be achieved by adopting such as dual-reference-layer structure[39,40] and ultralow-damping materials[41]. Because the *shape-anisotropy* MTJ can be formed with various structures using various material systems, there should remain a plenty of room of engineering to obtain better performance.

In summary, this study revisits the shape anisotropy, which has existed in conventional perpendicular MTJs but has so far counteracted the total perpendicular anisotropy because of its negative sign. We show that increasing the thickness of the recording layer makes the contribution positive and allows sufficient *Δ* of more than 80 at the 10-nm scale, which is not readily achieved by the conventional *interfacial-anisotropy* approach, as well as magnetization switching solely driven by current. The *shape-anisotropy* MTJ does not require any special materials, allowing us to choose from a broad range of material systems appropriate for the STT switching, which include large $M_S$ materials. Moreover, the concept can be extended to ultrafine MTJs with in-plane easy axes, which may be suitable for spin-orbit torque-induced switching devices[42]. The present results provide an additional insight for the miniaturization of MTJs and open up the next era of nano-magnetics/spintronics.

## Methods

**Parameters for calculation.** The analytical calculation of *Δ* and $J_{C0}$ shown in Fig. 1 were conducted under an assumption of typical double-interface MgO/FeB/MgO structures. The used parameters were as follows: $M_S$ = 1.5 T, $K_i$ = 2.0 mJ m$^{-2}$, $K_b$ = 0 J m$^{-3}$, $\alpha$ = 0.005 and TMR ratio = 100%. $g_{STT}$ was deduced from the relative angle of magnetizations across the MgO barrier layer and spin polarization via TMR according to the Jullière model[43].

**Film deposition.** The stacks were deposited by magnetron sputtering at room temperature onto 3-inch thermally-oxidized Si wafers. The stack structure was, from the substrate side, Ta(5)/ Ru(10)/ Ta(15)/ Pt(5)/ [Co(0.4)/ Pt(0.4)]$_{\times 6}$/ Co(0.4)/ Ru(0.4)/ Co(0.4)/ [Pt(0.4)/ Co(0.4)]$_{\times 2}$/ Ta(0.2)/ CoFeB(1)/ MgO(0.93)/ FeB(15)/ MgO(0.90)/ Ru(5) (nominal thickness in nm), where MgO layers were deposited by RF magnetron sputtering and the other layers were deposited by DC magnetron sputtering. The nominal composition of CoFeB was $(Co_{0.25}Fe_{0.75})_{75}B_{25}$, while that in FeB was $Fe_{75}B_{25}$ (at.%). The reference layer has a synthetic ferrimagnetic (SyF) structure, to minimize the shift of *R-H* curves. For the blanket film studies, where magnetic parameters were quantified, stacks with the recording layer part, namely, Ta(5)/ CoFeB(0.5)/ MgO(1.2)/ FeB($t$ = 2 - 30)/ MgO(1.2)/ CoFeB(0.5)/ Ta(1)/ Ru(5), were used.

**Nano-fabrication of MTJ.** The deposited film was processed into nano MTJs with electrodes. After the film deposition we deposited 80-nm Ta by sputtering *in-situ* and 50-nm Si-N by chemical vapor deposition *ex-situ*. Then we patterned the MTJs using conventional electron beam lithography. A Si-N hard mask was formed using the resist pattern, followed by a formation of Ta hard mask using the Si-N hard mask through a reactive ion etching. MTJs were defined by multistep Ar ion milling using the Ta-hard mask while varying the beam angle. During the milling, we monitored the secondary ion mass spectra and stopped the milling when the signal from the 10-nm-thick Ru layer appeared. The bottom Ta(5)/ Ru(10) layer was used as an electrode for the electrical measurement. After the milling, the MTJ was *ex-situ* covered with a Si-N passivation layer using chemical vapor deposition. This process might cause a formation of electrically dead regions through a side-wall oxidation [see Supplementary Note 3]. After that, the wafer was covered with a spin-on glass and then etched-back with a reactive ion etching until the tops of the Ta-hard mask were appeared. To form electrical contacts, Cr(5)/ Au(100) was deposited on top of the MTJs and Ta/Ru underlayer using photolithography and lift-off. The processed wafers were post-annealed in vacuum for 1 hour at 300°C.

**Measurements.** All the measurements were performed at room temperature. Vibrating sample magnetometer and ferromagnetic resonance using a TE$_{011}$ microwave cavity were used for the blanket film study, and a standard electrical probing system, capable of applying out-of-plane or in-plane dc magnetic fields, was used for measurement on nano MTJs. A current source was used to supply current pulses through the MTJ, and a voltmeter was used to measure the dc voltage across the MTJ. A dc current with typical magnitude of 0.5 or 1 μA was supplied when measuring the MTJ resistance.

**Data availability.** The data which support the findings of this work are available from the corresponding author upon reasonable request.

**References**

1  Slonczewski, J. C. Current-driven excitation of magnetic multilayers. *J. Magn. Magn. Mater.* **159**, L1-L7 (1996).
2  Berger, L. Emission of spin waves by a magnetic multilayer traversed by a current. *Phys. Rev. B* **54**, 9353-9358 (1996).
3  Tsoi, M. *et al.* Excitation of a Magnetic Multilayer by an Electric Current. *Phys. Rev. Lett.* **80**, 4281-4284 (1998).
4  Myers, E. B., Ralph, D. C., Katine, J. A., Louie, R. N. & Buhrman, R. A. Current-induced switching of domains in magnetic multilayer devices. *Science* **285**, 867-870 (1999).
5  Katine, J. A., Albert, F. J., Buhrman, R. A., Myers, E. B. & Ralph, D. C. Current-driven magnetization reversal and spin-wave excitations in Co /Cu /Co pillars. *Phys. Rev. Lett.* **84**, 3149-3152 (2000).
6  Huai, Y., Albert, F., Nguyen, P., Pakala, M. & Valet, T. Observation of spin-transfer switching in deep submicron-sized and low-resistance magnetic tunnel junctions. *Appl. Phys. Lett.* **84**, 3118-3120 (2004).
7  Diao, Z. *et al.* Spin transfer switching and spin polarization in magnetic tunnel junctions with MgO and AlOx barriers. *Appl. Phys. Lett.* **87**, 232502 (2005).
8  Kubota, H. *et al.* Magnetization switching by spin-polarized current in low-resistance magnetic tunnel junction with MgO [001] barrier. *IEEE Trans. Magn.* **41**, 2633-2635 (2005).
9  Hayakawa, J. *et al.* Current-driven magnetization switching in CoFeB/MgO/CoFeB magnetic tunnel junctions. *Jpn J Appl Phys 2* **44**, L1267-L1270 (2005).
10 Hosomi, M. *et al.* A novel nonvolatile memory with spin torque transfer magnetization switching: Spin-RAM. *IEDM Tech. Dig.*, 473-476 (2005).
11 Ralph, D. C. & Stiles, M. D. Spin transfer torques. *J. Magn. Magn. Mater.* **320**, 1190-1216 (2008).
12 Takemura, R. *et al.* A 32-Mb SPRAM With 2T1R Memory Cell, Localized Bi-Directional Write Driver and `1'/`0' Dual-Array Equalized Reference Scheme. *IEEE J. Solid-State Circuits* **45**, 869-879 (2010).

13 Brataas, A., Kent, A. D. & Ohno, H. Current-induced torques in magnetic materials. *Nature Mater.* **11**, 372-381 (2012).

14 Ohsawa, T. *et al.* A 1.5nsec/2.1nsec Random Read/Write Cycle 1Mb STT-RAM Using 6T2MTJ Cell with Background Write for Nonvolatile e-Memories. *Symp. VLSI Circuits Dig. Tech. Papers*, 110 (2013).

15 Kent, A. D. & Worledge, D. C. A new spin on magnetic memories. *Nature Nanotech.* **10**, 187-191 (2015).

16 Apalkov, D., Dieny, B. & Slaughter, J. M. Magnetoresistive Random Access Memory. *Proc. IEEE* **104**, 1796-1830 (2016).

17 Slaughter, J. M. *et al.* Technology for Reliable Spin-Torque MRAM Products. *IEDM Tech. Dig.*, 21.25.21-21.25.24 (2016).

18 Chung, S. W. *et al.* 4Gbit density STT-MRAM using perpendicular MTJ realized with compact cell structure. *IEDM Tech. Dig.*, 27.21.21-27.21.24 (2016).

19 Song, Y. J. *et al.* Highly functional and reliable 8Mb STT-MRAM embedded in 28nm logic. *IEDM Tech. Dig.*, 27.22.21-27.22.24 (2016).

20 Parkin, S. S. *et al.* Giant tunnelling magnetoresistance at room temperature with MgO (100) tunnel barriers. *Nature Mater.* **3**, 862-867 (2004).

21 Yuasa, S., Nagahama, T., Fukushima, A., Suzuki, Y. & Ando, K. Giant room-temperature magnetoresistance in single-crystal Fe/MgO/Fe magnetic tunnel junctions. *Nature Mater.* **3**, 868-871 (2004).

22 Mangin, S. *et al.* Current-induced magnetization reversal in nanopillars with perpendicular anisotropy. *Nature Mater.* **5**, 210-215 (2006).

23 Meng, H. & Wang, J.-P. Spin transfer in nanomagnetic devices with perpendicular anisotropy. *Appl. Phys. Lett.* **88**, 172506 (2006).

24 Seki, T., Mitani, S., Yakushiji, K. & Takanashi, K. Spin-polarized current-induced magnetization reversal in perpendicularly magnetized L10-FePt layers. *Appl. Phys. Lett.* **88**, 172504 (2006).

25 Nakayama, M. *et al.* Spin transfer switching in TbCoFe/CoFeB/MgO/CoFeB/TbCoFe magnetic tunnel junctions with perpendicular magnetic anisotropy. *J. Appl. Phys.* **103**, 07A710 (2008).

26 Hayakawa, J. *et al.* Current-Induced Magnetization Switching in MgO Barrier Magnetic Tunnel Junctions With CoFeB-Based Synthetic Ferrimagnetic Free Layers. *IEEE Trans. Magn.* **44**, 1962-1967 (2008).

27 Yakata, S. *et al.* Influence of perpendicular magnetic anisotropy on spin-transfer switching current in CoFeB/MgO/CoFeB magnetic tunnel junctions. *J. Appl. Phys.* **105**, 07D131 (2009).

28 Ikeda, S. *et al.* A perpendicular-anisotropy CoFeB-MgO magnetic tunnel junction. *Nature Mater.* **9**, 721-724 (2010).

29 Worledge, D. C. *et al.* Spin torque switching of perpendicular Ta│CoFeB│MgO-based magnetic tunnel junctions. *Appl. Phys. Lett.* **98**, 022501 (2011).

30 Sun, J. Z. *et al.* Spin-torque switching efficiency in CoFeB-MgO based tunnel junctions. *Phys. Rev. B* **88** (2013).

31 Kubota, H. *et al.* Enhancement of perpendicular magnetic anisotropy in FeB free layers using a thin MgO cap layer. *J. Appl. Phys.* **111**, 07C723 (2012).

32 Sato, H. *et al.* Perpendicular-anisotropy CoFeB-MgO magnetic tunnel junctions with a MgO/CoFeB/Ta/CoFeB/MgO recording structure. *Appl. Phys. Lett.* **101**, 022414 (2012).

33 Jan, G. *et al.* High Spin Torque Efficiency of Magnetic Tunnel Junctions with MgO/CoFeB/MgO Free Layer. *Appl. Phys. Express* **5**, 093008 (2012).

34 Sato, H. *et al.* Properties of magnetic tunnel junctions with a MgO/CoFeB/Ta/CoFeB/MgO recording structure down to junction diameter of 11 nm. *Appl. Phys. Lett.* **105**, 062403 (2014).

35 Hallal, A., Yang, H. X., Dieny, B. & Chshiev, M. Anatomy of perpendicular magnetic anisotropy in Fe/MgO magnetic tunnel junctions: First-principles insight. *Phys. Rev. B* **88** (2013).

36 Peng, S. *et al.* Interfacial Perpendicular Magnetic Anisotropy in Sub-20 nm Tunnel Junctions for Large-Capacity Spin-Transfer Torque Magnetic Random-Access Memory. *IEEE Magn. Lett.* **8**, 1-5 (2017).

37 Maehara, H. *et al.* Tunnel Magnetoresistance above 170% and Resistance–Area Product of 1 Ω (μm)2 Attained by In situ Annealing of Ultra-Thin MgO Tunnel Barrier. *Appl. Phys. Express* **4**, 033002 (2011).

38 Koch, R. H., Katine, J. A. & Sun, J. Z. Time-resolved reversal of spin-transfer switching in a nanomagnet. *Phys. Rev. Lett.* **92**, 088302 (2004).

39 Diao, Z. *et al.* Spin transfer switching in dual MgO magnetic tunnel junctions. *Appl. Phys. Lett.* **90**, 132508 (2007).

40 Hu, G. *et al.* STT-MRAM with double magnetic tunnel junctions. 26.23.21-26.23.24 (2015).

41 Schoen, M. A. W. *et al.* Ultra-low magnetic damping of a metallic ferromagnet. *Nature Physics* **12**, 839-842 (2016).

42 Fukami, S., Anekawa, T., Zhang, C. & Ohno, H. A spin-orbit torque switching scheme with collinear magnetic easy axis and current configuration. *Nature Nanotech.* **11**, 621-625 (2016).

43 Jullière, M. Tunneling between Ferromagnetic-Films. *Phys. Lett. A* **54**, 225-226 (1975).

**Acknowledgements**

The authors thank S. Ikeda, F. Matsukura and J. Llandro for fruitful discussion and C. Igarashi, T. Hirata, H. Iwanuma, Y. Kawato, K. Goto, I. Morita, R. Ono, and M. Musya for their technical support. A portion of this work was supported by R&D Project for ICT Key Technology to Realize Future Society of MEXT, JST-OPERA, ImPACT Program of CSTI, JSPS KAKENHI Grant Numbers 17H06093, 15H05521, and Cooperative Research Projects of RIEC. K.W. acknowledges the Graduate Program in Spintronics, Tohoku University.

**Author contributions**

K.W., S.F., H.S. and H.O. planned the study. K.W. and H.S. deposited the films, and K.W., B.J. and H.S. processed into devices. K.W. conducted the measurements, analyzed the data and performed the calculation. K.W. and S.F. wrote the manuscript with input from B.J., H.S. and H.O. All authors discussed the results.

**Competing interests**

The authors declare no competing financial interests.

## Supplementary Note 1. Shape anisotropy coefficient δN

The demagnetization coefficient is the ratio of the magnetostatic energy increase when the magnet considered is uniformly magnetized along arbitrary directions. In cartesian coordinate system, the coefficients $N_x$, $N_y$ and $N_z$ satisfy

$$N_x + N_y + N_z = 1. \tag{1}$$

When $z$ axis is along the film normal direction and the magnet is isotropic in the film plane, *i.e.*, circular shape, $x$ and $y$ axes are equivalent and thus

$$N_x = N_y. \tag{2}$$

Since the magnetic anisotropy is defined as the difference in energy density between the states where magnetization aligns in the perpendicular and in-plane direction, shape anisotropy coefficient δN is given by

$$\delta N = N_z - N_x. \tag{3}$$

To analytically calculate the demagnetization coefficients of the recording layer with a cylindrical shape, we approximate it by a spheroidal single-domain nanomagnet with an aspect ratio $q = t/D$. Then, the demagnetization coefficient is known to be given by the following equations for two cases:

- Oblate spheroid $(q < 1)$[1],

$$N_x = \frac{q^2}{2(1-q^2)}\left(\frac{1}{q\sqrt{1-q^2}}\cos^{-1} q - 1\right). \tag{4}$$

- Prolate spheroid $(q \geq 1)$[2],

$$N_z = \frac{1}{q^2}\left\{\frac{1}{3} + \sum_{k=1}^{\infty}\frac{1}{2k+3}\left(1-\frac{1}{q^2}\right)^k\right\}. \tag{5}$$

The calculated δN from Supplementary Equations 3, 4, and 5 is shown in Supplementary Figure 1a. The two curves with different approximations are connected smoothly with each other. Thus, we use Supplementary Equation 4 for $q \leq 0.9$ and Supplementary Equation 5 for $q > 0.9$ to calculate δN in equation (1) of the main body. Supplementary Figure 1b shows $t$ and $D$ dependence of δ$N$ calculated from Supplementary Figure 1a. One can see that increasing $t$ with $D$ is required to satisfy the condition δ$N$ > 1, resulting in a requirement of increasing $t$ with $D$ for larger $D$ region in Fig. 1a. On the other hand, for smaller $D$ region, increasing $t$ with decreasing $D$ is necessary to keep the volume of the free layer.

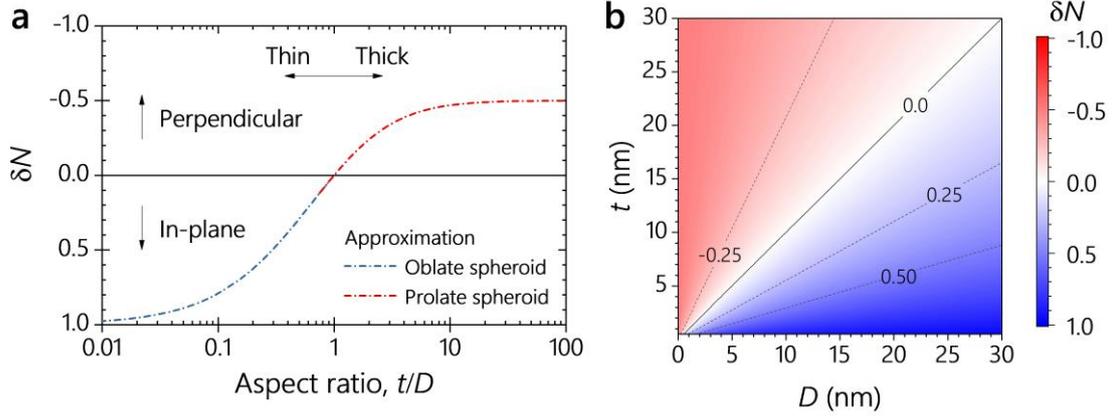

**Supplementary Figure 1 | Calculated shape anisotropy coefficient δ$N$. a**, Aspect ratio $t$/$D$ dependence of δ$N$ calculated according to the two approximations[1,2]. **b**, δ$N$ as functions of $t$ and $D$.

## Supplementary Note 2. Magnetic properties of MgO/FeB/MgO blanket films

The magnetic properties of MgO/FeB/MgO stacks are evaluated with the vibrating sample magnetometer (VSM) and ferromagnetic resonance (FMR) measurements for blanket films. Supplementary Figures 2a and 2b, respectively, are the areal magnetic moment $m$ in response to in-plane and out-of-plane magnetic field $H$ ($m$-$H$ curves). Saturated value of $m$ ($m_S$) and areal effective magnetic anisotropy energy density $K_{eff}t$, obtained from the difference in the in-plane and out-of-plane $m$-$H$ curves, are plotted as a function of FeB thickness $t$ in Supplementary Figures 2c and 2d. We determine the spontaneous magnetization $M_S$, magnetic dead layer thickness $t_d$, interfacial and bulk magnetic anisotropy energy density $K_i$ and $K_b$ by the same way as used in Supplementary Reference 3. The obtained values are $M_S$ = 1.52 ± 0.01 T, $t_d$ = 0.0 ± 0.1 nm, $K_i$ = 2.2 ± 0.1 mJ m$^{-2}$, $K_b$ = (− 1.10 ± 0.07) ×10$^5$ J m$^{-3}$.

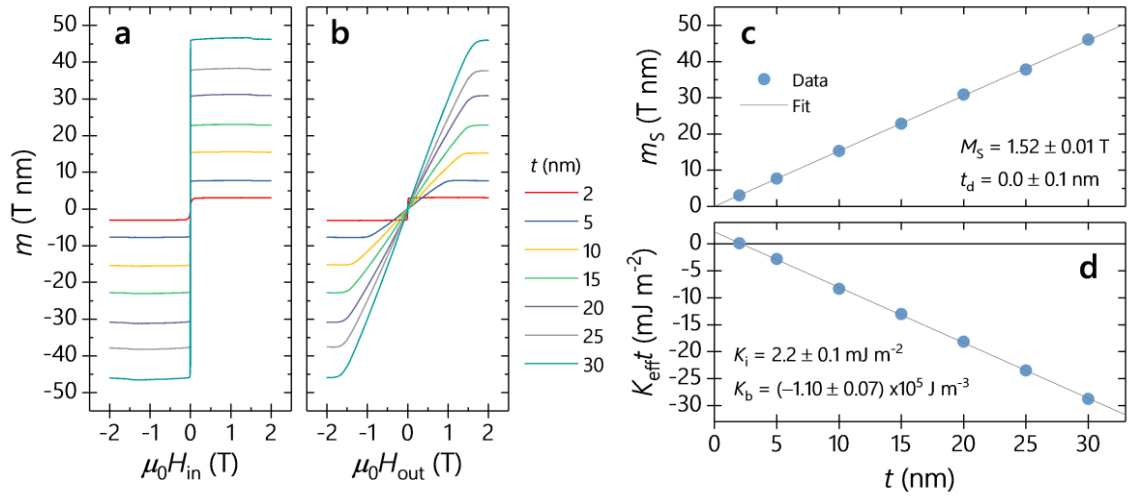

**Supplementary Figure 2 | Magnetic properties characterization by VSM. a,b**, Areal magnetic moment *m* versus applied magnetic field *H* (*m-H*) curves with various FeB thickness *t*. **a**, *m-H* curves along in-plane direction. **b**, *m-H* curves along out-of-plane direction. **c,d**, *t* dependence of **c**, $m_S$ and **d**, $K_{eff}t$ as a function of *t* with linear fitting.

Supplementary Figure 3a shows the FMR spectra measured at various magnetic field angles $\theta_H$ from the film normal direction on a blanket film with $t$ = 20 nm. Resonance field $H_R$ and full-width at half-maximum $\Delta H$ of each spectrum is extracted and plotted in Supplementary Figures 3b and 3c, respectively, as a function of $\theta_H$. The dependence of $\Delta H$ on $\theta_H$ is fitted with the same procedure including extrinsic effects such as two-magnon scattering and motional narrowing as that in Supplementary Reference 4. $\alpha$ is determined to be 0.00425 ± 0.00003.

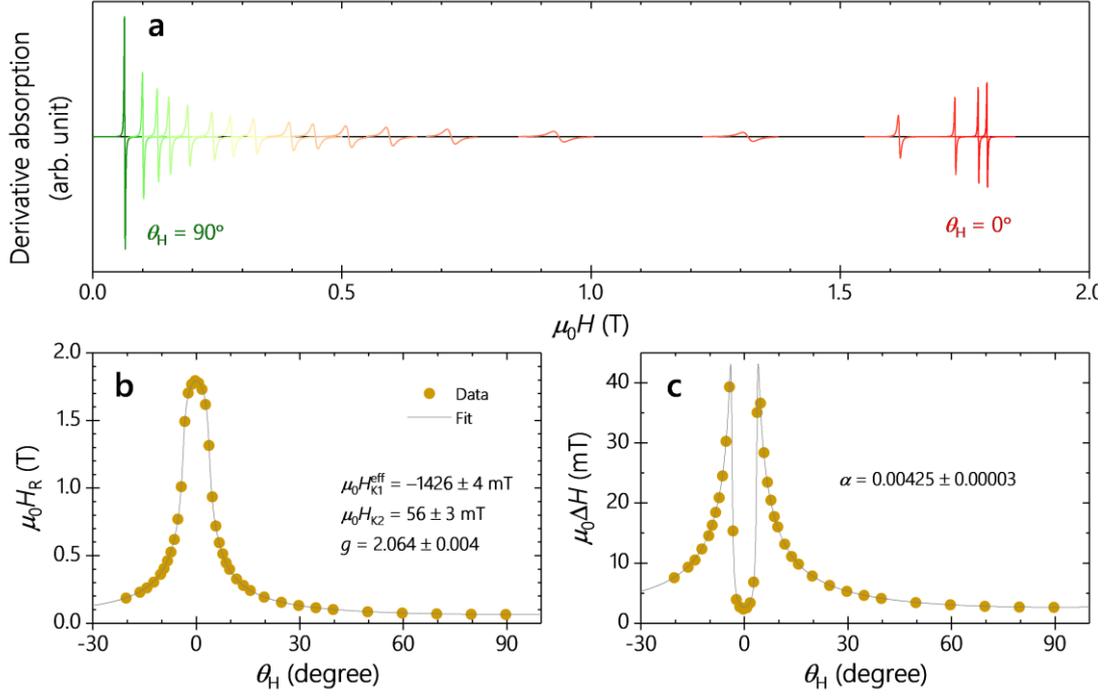

**Supplementary Figure 3 | Magnetic property characterization by FMR measurement. a**, FMR spectra measured at various magnetic field angles $\theta_H$. **b**, $\theta_H$ dependence of resonance field $H_R$ and its fitting. **c**, $\theta_H$ dependence of full-width at half-maximum $\Delta H$ and its fitting.

## Supplementary Note 3. Resistance-area product *RA*

The square root of the electrically-measured values of conductance *G* at low resistive state of MTJs are plotted as a function of the physical diameter $D_{TEM}$ of MgO barrier layer observed by transmission electron microscopy in Supplementary Figure 4. The resistance-area product *RA* is determined by a linear fitting according to Supplementary Equation 6, in which an electrically dead region[5] with a length of $D_0$ that surrounds electrically active region with diameter *D* is considered as a horizontal intercept:

$$G = \frac{1}{RA}\frac{\pi(D_{TEM} - D_0)^2}{4}. \tag{6}$$

*RA* of our MTJs is determined from the fitting to be $4.5 \pm 0.5\ \Omega\ \mu m^2$, with the depth of electrically dead region ($D_0/2$) to be $2.5 \pm 0.7$ nm.

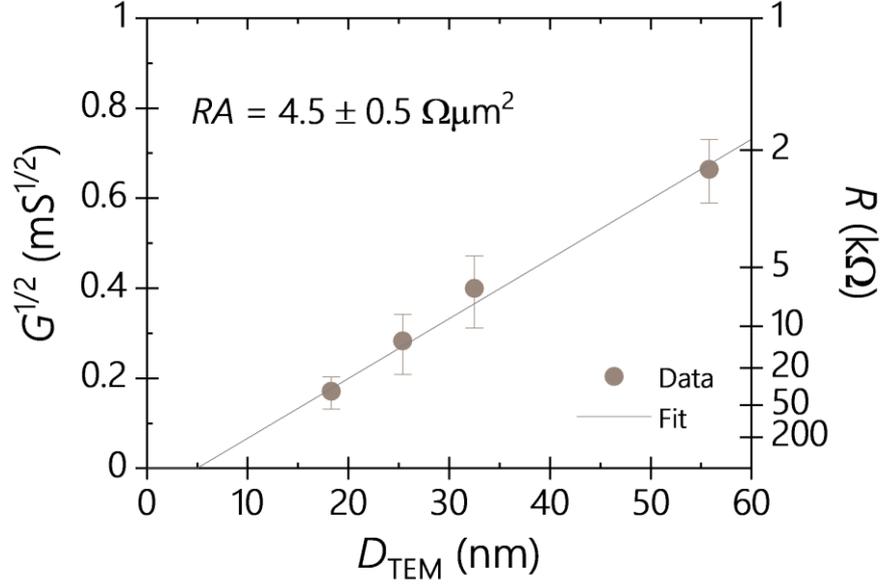

**Supplementary Figure 4 | Determination of resistance-area product *RA*.** The square root of the MTJ conductance *G* as a function of MgO diameter $D_{\text{TEM}}$ determined from TEM observation. The solid line is a fitting according to Supplementary Equation 6. The right axis shows the correspondent resistance *R*.

## Supplementary Note 4. Switching probability measurement using pulse magnetic field

In order to quantify the thermal stability factor $\Delta$ of nano MTJs, switching probability measurement is carried out using pulse magnetic field with its duration $\tau$ of 1 s. The probability of magnetization reversal based on the Stoner-Wohlfarth model under an application of magnetic field is described as[6]

$$P_{\text{P(AP)}} = 1 - \exp\left[-\frac{\tau}{\tau_0}\exp\left\{-\Delta\left(1 \mp \frac{H - H_{\text{shift}}}{H_{\text{K}}^{\text{eff}}}\right)^2\right\}\right], \tag{7}$$

where subscript P(AP) designates the magnetization switching from parallel (anti-parallel) to anti-parallel (parallel). $\tau_0$ is the inverse of attempt frequency assumed to be 1 ns, $H_{\text{shift}}$ is the shift field and $H_{\text{K}}^{\text{eff}}$ is the effective magnetic anisotropy field. Supplementary Figures 5a and 5b show the *R-H* loop and the switching probability as a function of the amplitude of magnetic-field pulse for an MTJ with $D = 7.3 \pm 0.4$ nm. The result of the switching probability is fitted by Supplementary Equation 7 with $\Delta$, $H_K^{eff}$ and $H_{shift}$ as fitting parameters, leading to $\Delta = 91 \pm 2$, $\mu_0 H_K^{eff} = 207 \pm 2$ mT and $\mu_0 H_{shift} = 10.2 \pm 0.1$ mT.

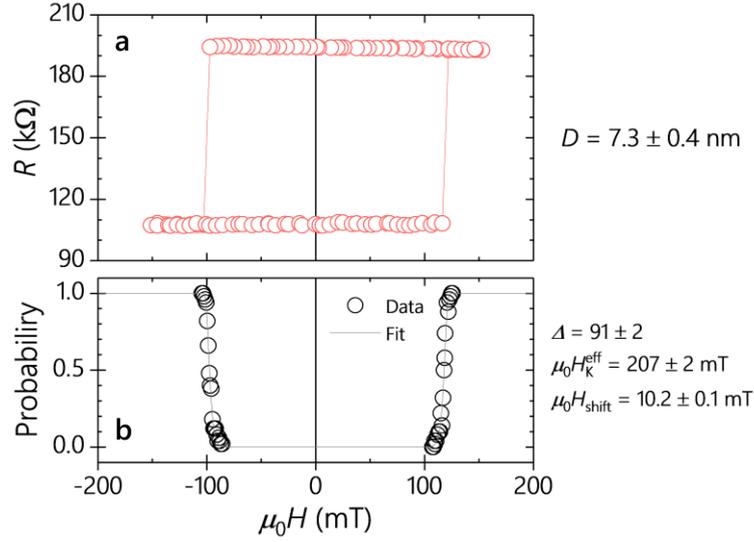

**Supplementary Figure 5 | *R-H* loop and determination of *Δ*. a**, an *R-H* loop measured for an MTJ with *D* = 7.3 nm. **b**, Switching probability as a function of 1 s-long pulse magnetic field and the fitting according to Supplementary Equation 7.

## Supplementary References

1   Osborn, J. A. Demagnetizing Factors of the General Ellipsoid. *Phys. Rev.* **67**, 351-357 (1945).

2   Aharoni, A. *Introduction to the Theory of Ferromagnetism*. (Oxford University Press Inc., 1996).

3   Watanabe, K. *et al.* Annealing temperature dependence of magnetic properties of CoFeB/MgO stacks on different buffer layers. *Jpn. J. Appl. Phys.* **56**, 0802B0802 (2017).

4   Okada, A. *et al.* Magnetization dynamics and its scattering mechanism in thin CoFeB films with interfacial anisotropy. *Proc. Natl. Acad. Sci. USA* **114**, 3815-3820 (2017).

5   Gajek, M. *et al.* Spin torque switching of 20 nm magnetic tunnel junctions with perpendicular anisotropy. *Appl. Phys. Lett.* **100**, 132408 (2012).

6   Li, Z. & Zhang, S. Thermally assisted magnetization reversal in the presence of a spin-transfer torque. *Phys. Rev. B* **69** (2004).